\documentclass[
]{ceurart}

\usepackage{listings}
\begin{document}

\copyrightyear{2022}
\copyrightclause{Copyright for this paper by its authors.
  Use permitted under Creative Commons License Attribution 4.0
  International (CC BY 4.0).}

\conference{Woodstock'22: Symposium on the irreproducible science,
  June 07--11, 2022, Woodstock, NY}

\title{Approximate-At-Most-k Encoding of SAT \\ for Soft Constraints}

\author[1]{Shunji Nishimura}[%
orcid=0000-0001-6600-5136,
email=s-nishimura@oita-ct.ac.jp,
url=https://onct.oita-ct.ac.jp/seigyo/nishimura_hp/,
]
\address[1]{National Institute of Technology, Oita College,
  1666 Maki, Oita City, Oita Prefecture, Japan}

\begin{abstract}
In the field of Boolean satisfiability problems (SAT), at-most-$k$ constraints, which suppress the number of true target variables at most $k$, are often used to describe objective problems.
At-most-k constraints are used not only for absolutely necessary constraints (hard constraints) but also for challenging constraints (soft constraints) to search for better solutions.
To encode at-most-k constraints into Boolean expressions, there is a problem that the number of Boolean expressions basically increases exponentially with the number of target variables, so at-most-k often has difficulties for a large number of variables.
To solve this problem, this paper proposes a new encoding method of at-most-k constraints, called approximate-at-most-k, which has totally fewer Boolean expressions than conventional methods on the one hand.
On the other hand, it has lost completeness, i.e., some Boolean value assignments that satisfy the original at-most-k are not allowed with approximate-at-most-k; hence, it is called approximate.
Without completeness, we still have potential benefits by using them only as soft constraints.
For example, approximate-at-most-16 out of 32 variables requires only 15\% of a conventional at-most-k on the literal number and covers 44\% of the solution space.
Thus, approximate-at-most-k can become an alternative encoding method for at-most-k, especially as soft constraints.
\end{abstract}



\maketitle

\section{Introduction}

SAT, or the Boolean satisfiability problem, demonstrates its availability for real-world problems in many areas.
To tackle a real-world problem, we have to describe it as a combination of constraints compatible with SAT, and there is a commonly used constraint called \textit{at-most-$k$}, along with some Boolean variables, which is satisfied if the variables have at most $k$ number of trues in total.

One of the problems with at-most-k constraints is the combinatorial explosion; the number of encoded Boolean expressions for an at-most-k constraint will explode along with increasing the target variables.
To alleviate the restriction around the explosion problem, several encodings \cite{frisch2010sat,bittner2019sat} have been proposed such as binary encoding, sequential counter encoding, commander encoding, product encoding, etc.
While all of these are genuinely at-most-k constraints, of course, this paper provides a different approach to the problem that attempts to drastically reduce the number of Boolean expressions, in exchange for losing an accurate count of trues.
The encoding method we propose, called \textit{approximate-at-most-k}, is no longer genuine at-most-k because some parts of solutions for the original at-most-k may not be included in the solution space of our approximate-at-most-k.
For example, assignment $(X_1,X_2,X_3,X_4,X_5)=(true,true,false,false,true)$ has three trues, so satisfies at-most-3, but may not be satisfied with approximate-at-most-3, depending on the model implemented at that time.
In terms of proof theory, where the judgment of a SAT solver is regarded as the existence of proof, we can say that approximate-at-most loses \textit{completeness}.

Despite the lack of completeness, approximate-at-most can still be useful in some cases, such as using them as \textit{soft constraints} \cite{meseguer2006soft,schiendorfer2013constraint}, or preferences, in other words, that can be used to describe optional desires around the objective problem.
For an example of university timetabling \cite{bettinelli2015overview,bittner2019sat}, on the one hand, it is necessary that the same teacher not teach two different classes at the same time (this is called a hard constraint).
On the other hand, a university policy is suitable for soft constraints; it may be preferable that only 5 teachers have continuous classes, rather than 10 teachers.
For soft constraints, we assume that it is not necessary to exactly evaluate satisfiability, and we can compare the benefit of the solution coverage with the cost of their Boolean expressions.

The fundamental idea of approximate-at-most-k is common to fractional encoding \cite{ourkris} .
While fractional encoding has completeness, approximate-at-most-k does not, as mentioned above, and focuses only on reducing the number of Boolean expressions.

\section{Approximate At-Most-k Encoding}

\subsection{Fundamental idea}
An example is shown in Fig. \ref{fig:FundamentalIdea}, which illustrates the idea of approximate-at-most-k encoding.
First, set $A$ of four variables (depicted as circles) must be constrained by at-most-2.
Next, the number of trues in $A_1$, the left half of $A$, constrains variables $B_1$, the left half of the set of variables $B$.
Specifically, as follows:
\begin{itemize}
    \item when 0 trues in $A_1$, $B_1$ is constrained by at-most-0,
    \item when 1 true in $A_1$, $B_1$ is constrained by at-most-2,
    \item when 2 trues in $A_1$, $B_1$ is constrained by at-most-4 (makes no sense).
\end{itemize}
In general, $B_1$ is dynamically constrained by the number of trues in $A_1$.
Since at most two in $A$ can be true, at most four in $B$ can be true.
Thus, these constraints in total behave as an at-most-4 constraint on $B$.

Note that this is not a proper at-most-4 constraint because some cases of possible solutions are missing.
For example, if $B_1$ has three trues and one true in the other variables of $B$, the right half, then that case satisfies an at-most-4 constraint on $B$ but does not satisfy our idea given above.
Actually, in that case, $A_1$ needs two trues and the right half of $A$ needs (at least) one true, and that is not possible under at-most-2 constraint on $A$.
Because of this incompleteness, we call the idea \textit{approximate-at-most}.
We have to be careful about approximate-at-most not to use for determination of satisfiability, but to use only for searching better solutions along with soft constraints.

That approximate-at-most-4-of-8 constraint is composed of a few at-most-2-of-4 constraints.
Since the number of Boolean expressions for at-most constraints grows exponentially with the size of target variables, roughly speaking, we may expect the number of Boolean expressions on approximate-at-most-k will be reduced in some cases, as it were ''single large or several small.``

\subsection{$2 \times 2$ models}
\begin{figure}[t]
 \centering
 \begin{minipage}{0.49\textwidth}
  \centering
  \includegraphics[width=90mm]{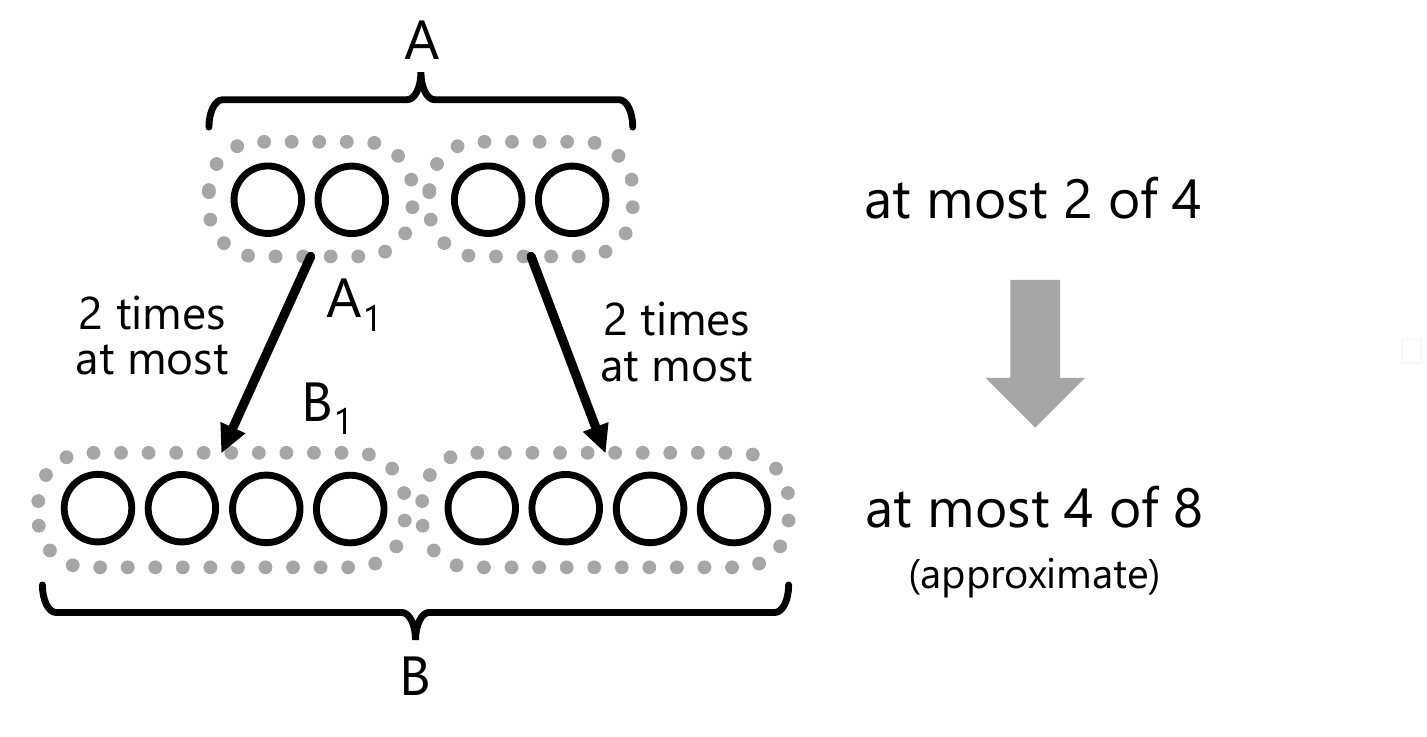}
  \caption{The fundamental idea to transform \\ at-most-2 on variables in $A$ into approximate \\ at-most-4 on variables in $B$.}
  \label{fig:FundamentalIdea}
 \end{minipage}
 \hfil
  \begin{minipage}{0.49\textwidth}
  \includegraphics[width=50mm]{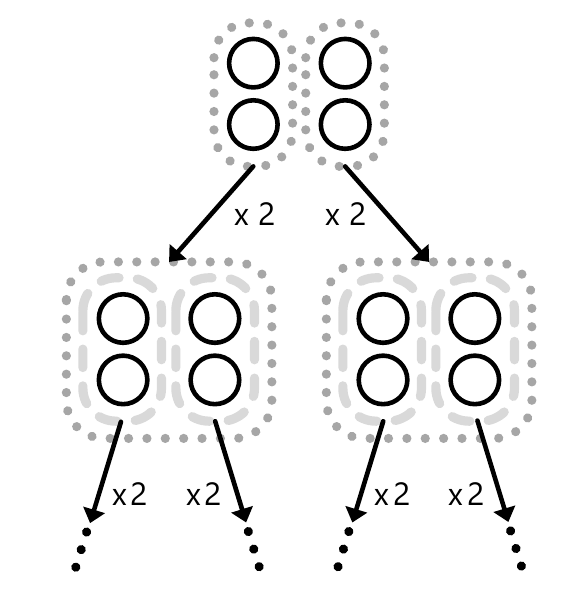}
  \caption{$2 \times 2$ models are recursively defined \\ between two parent variables and four child \\variables.}
  \label{fig:2x2}
 \end{minipage}
\end{figure}

The idea is able to apply to tree structure recursively, as shown in Fig. \ref{fig:2x2}.
Two Boolean variables in the same column at a parent node constrain corresponding four Boolean variables at the child node;
 when there are number $n \, (0 \leq n \leq 2)$ of trues in the two variables of the parent, the four child variables are constrained by at-most-$2n$.
In Boolean expressions,
\begin{equation}
\neg\, v_1 \Rightarrow AtMost_0\{u_{11},u_{12},u_{21},u_{22}\} 
\quad\wedge\quad
\neg\, v_2 \Rightarrow AtMost_2\{u_{11},u_{12},u_{21},u_{22}\}
\label{eq:2x2}
\end{equation}
where $v_i$ and $u_{ij}$ denote variables in a parent node and child node respectively, and we assume $v_i$ are in order encoding \cite{tamura2009compiling}, i.e., $v_2 \Rightarrow v_1$ holds.
By giving at-most-$k$ $(0 \leq k \leq 4)$ at the four variables of the top, these models generate approximate-at-most-$(k/4 \cdot 2^{m+1})$ of $2^{m+1}$ at the bottom, where $m=1,2,\cdots$ denotes the height of  the tree.

\subsection{generalized $h \times w$ models}
\begin{figure}
    \centering
    \includegraphics[width=150mm]{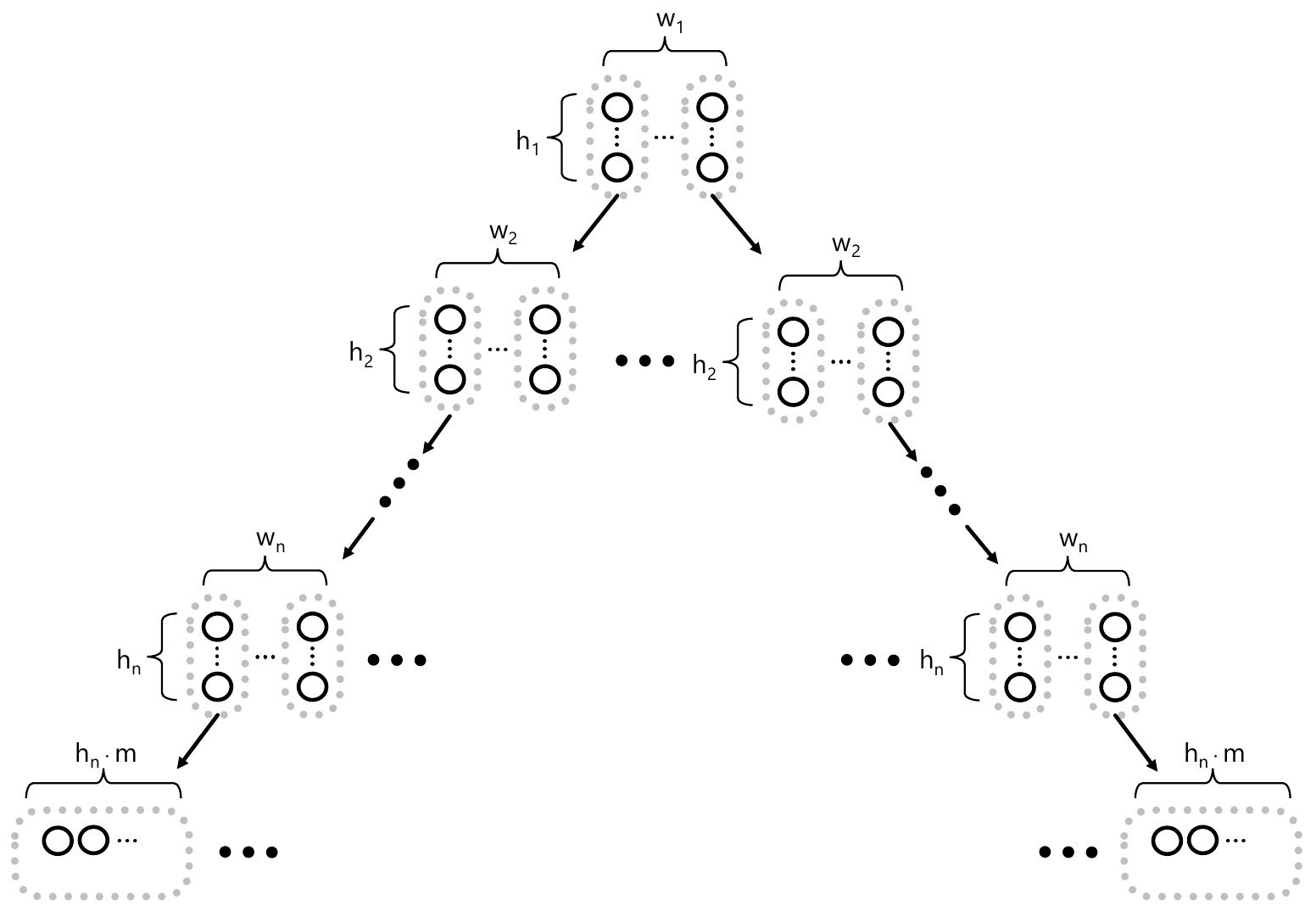}
    \caption{$h \times w$ models have arbitrary (but the same at the same level) numbers of height and width.}
    \label{fig:hxw}
\end{figure}

More generalized models are shown in Fig. \ref{fig:hxw}, in which each node except the bottoms has a matrix of variables with height $h$ and width $w$.
On the same hierarchy level, the height and width of nodes are identical.
Between a parent column of height $h_i$ and its children of $h_{i+1} \times w_{i+1}$, we need $h_{i+1} \cdot w_{i+1}$ to be multiple of $h_i$, i.e. $h_{i+1} \cdot w_{i+1} \,mod\, h_i = 0$; when $h_{i+1} \cdot w_{i+1} = a \cdot h_i$ for some $a$ and $n$ trues in the parent column, the child variables are constrained by  at-most-$(a \cdot n)$.
In Boolean expressions,
\begin{equation}
    \bigwedge\limits_{j=1,\cdots,h} \neg\, v_j \Rightarrow AtMost_{h' \cdot w' \cdot (j-1)/h} \{\text{child variables of $v$}\}
\end{equation}
where $v_j$ denotes parent variables of height $h$ and the child node has $h' \cdot w'$ variables. We also assume $v_j$  are in order encoding, i.e., $v_j \Rightarrow v_{j-1} (j=2,\cdots,h)$ holds.
For leaf nodes at the bottom, they are simply defined sets of variables of number $h_n \times m$, where $h_n$ is the parent's height and an arbitrary $m$.
By giving at-most-$k$ at the top node, these models generate approximate-at-most-$(k/(h_1 \cdot w_1) \cdot \Pi w_i \cdot h_n \cdot m)$-of-$(\Pi w_i \cdot h_n \cdot m)$.

For the sake of ease, let us also use a fraction  to denote the number of the constraints as approximate-at-most-$a/b$-of-$n$, when the top node has $b$ variables and at-most-$a$ is given, to generate approximate-at-most-$(a/b \cdot n)$-of-$n$ in the integer expression.

\section{Experimental Results}

All software materials for these experiments are on the GitHub repository \cite{TheRepos}.
\subsection{$2 \times 2$ models}
Here are the results of the number of Boolean expressions and coverages of the solution space, about $2 \times 2$ models.
The CNF (Conjunctive Normal Form) of approximate-at-most-1/2-of-16 shows the following:
\begin{itemize}
    \item auxiliary variables: 12
    \item clauses: 58
    \item literals: 168
\end{itemize}
where every two variables in columns are encoded by order encoding \cite{tamura2009compiling} and each at-most-k constraint for small numbers employs binomial (pairwise) encoding.
There are 39,203 possible solutions and approximate-at-most covers 68.2\% of them, overall, which means fewer numbers are included such as 7$\sim$0 true(s).
For the solutions of just 8 true variables, there are 12,870 possible solutions, and approximate-at-most covers 38.1\% of them.
While it will be depended on the objectives of using SAT whether to be focused on overall possible solutions or possible solutions of the maximal number, this paper mainly deals with the former, overall possible solutions from this point.

\begin{figure}[htb]
    \centering
    \includegraphics[width=110mm]{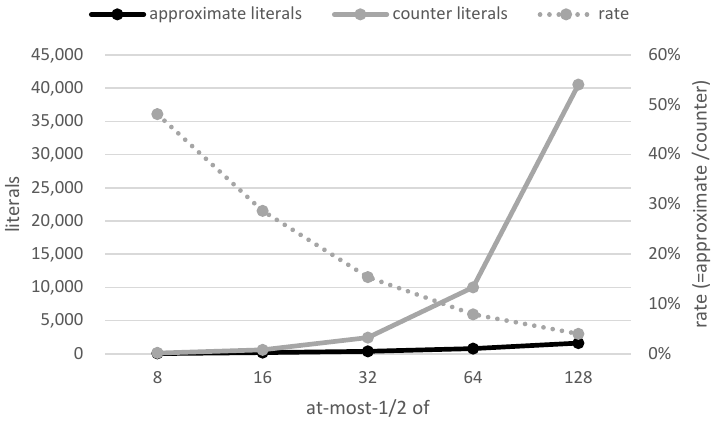}
    \caption{Literal numbers of approximate ($2 \times 2$ models) and counter at-most-1/2. As the number of target variables increases, the difference between approximate and counter is enlarged, and thus the rate of them, the dashed line, decreases.}
    \label{fig:Graph2x2Literals}
\end{figure}

Comparing approximate-at-most to conventional encoding methods, we focused on the literal number and choose counter encoding \cite{sinz2005towards,frisch2010sat}, that demonstrates superior results at the literal number among the other encodings: binomial, binary, commander, and product \cite{frisch2010sat}.
Counter encoding of at-most-$k$ is as follows.
\begin{align}
    & \bigwedge \limits_{i=1}^{n-1} \neg\, x_i \vee r_{i,1} \\
    & \bigwedge \limits_{j=2}^x \neg\, r_{1,j} \\
    & \bigwedge \limits_{i=2}^{n-1} \bigwedge \limits_{j=1}^k \neg\, r_{i-1,j} \vee r_{i,j} \\
    & \bigwedge \limits_{i=2}^{n-1} \bigwedge \limits_{j=2}^{k} \neg\, x_i \vee \neg\, r_{i-1,j-1} \vee r_{i,j} \\
    & \bigwedge \limits_{i=2}^n \neg\, x_i \vee \neg\, r_{i-1,k}
\end{align}
where $x_1,\cdots,x_n$ are target variables and $r_{i,j}$ denotes auxiliary variables.
The black and gray lines in  Fig. \ref{fig:Graph2x2Literals} indicate the literal number of approximate and counter at-most-1/2 respectively.
Literal rate is defined as below:
\begin{equation}
    \text{literal rate} = \text{(approximate literals)} \,/\, \text{(counter literals)}.
\end{equation}
As expected, approximate consumes a lower number than a conventional encoding.

\begin{figure}[tb]
    \centering
    \includegraphics[width=110mm]{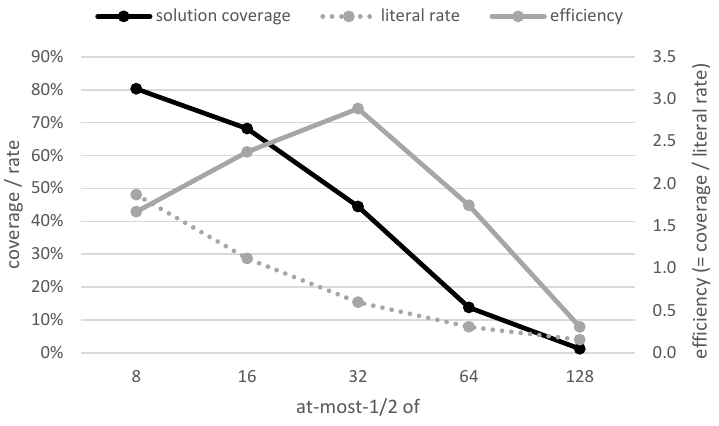}
    \caption{Coverages and efficiencies of approximate-at-most-1/2. Efficiency is defined as coverage par literal rate.}
    \label{fig:Graph2x2Efficiency}
\end{figure}

About the solution coverages, defined as below,
\begin{equation}
    \text{coverage} = \text{(solutions by approximate)} / \text{(all solutions)},
\end{equation}
the black line in Fig. \ref{fig:Graph2x2Efficiency}  indicates them. Predictably, it becomes lower as target variables increase.
With the literal rate of approximate/counter, the dashed line (same as in Fig. \ref{fig:Graph2x2Literals}), the coverage is larger than the literal rate at 8$\sim$64 variables but it turns around at 128 variables.
Since we naturally hope to gain maximal coverage by fewer literals, the value of coverage per literal rate, the gray line, could be regarded as \textit{efficiency} from that point of view, i.e.,
\begin{equation}
        \text{efficiency} = \text{coverage} \,/\, \text{(literal rate)}.
\end{equation}
In other words, the efficiency indicates a kind of advantage over counter encoding, with consideration for solution coverage.
In the range of Fig. \ref{fig:Graph2x2Efficiency} , approximate-at-most-1/2 of 32 variables exercises the most efficiently: 44.5\% coverage on 15.4\% literal rate (approximate: 376 / counter: 2,449).
\subsection{$h \times w$ models}
For an arbitrary $k$ and $n$, where $k < n$, there are many $h \times w$ models to implement approximate-at-most-$k$ of $n$ in general.
For an example of approximate-at-most-5 of 10, the followings are available:
\begin{itemize}
    \item $h_1=2,w_1=2,m=3$, at-most-2 on the top, fix 1 false and 1 true on the bottom variables
    \item $h_1=2,w_1=2,h_2=2,w_2=2,m=2$, at-most-2 on the top, fix 3 falses and 3 trues on the bottom variables
\end{itemize}
where $h_i$,$w_i$, and $m$ are as shown in Fig. \ref{fig:hxw}, and there are 8 other models.

\begin{figure}[tb]
    \centering
    \includegraphics[width=130mm]{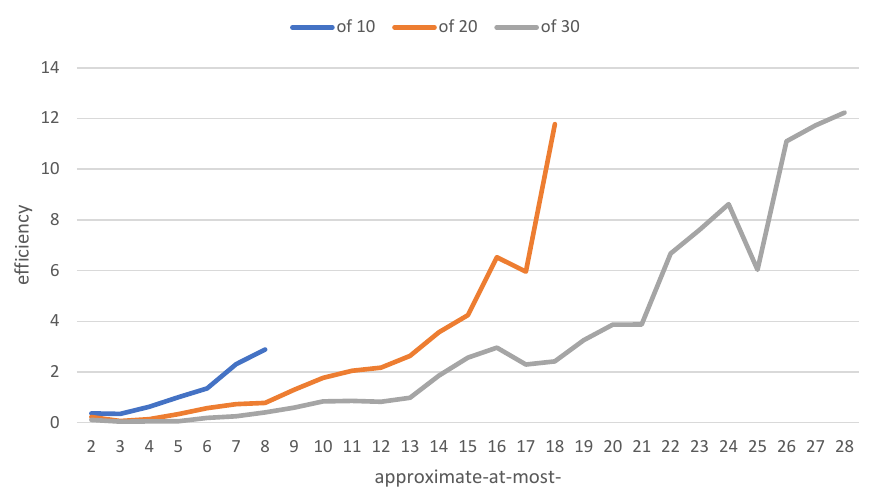}
    \caption{The best efficiencies (= coverage / literal rate vs counter) for each approximate-at-most-$k$,  of 10, 20, and 30.}
    \label{fig:TheBestEfficiencies}
\end{figure}

Among such models to implement approximate-at-most-$k$, we are naturally interested in the most efficient model, and Fig. \ref{fig:TheBestEfficiencies} shows the best efficiencies (= coverage / literal rate vs counter) for each approximate-at-most-$k$,  of 10, 20, and 30.
For instance, approximate-at-most-5 of 10 has the best efficiency on the model:
\[
h_1=2,w_1=3,m=2 \text{, at-most-3 on the top, fix 1 false and 1 true on the bottom variables}
\]
and shows the following.
\begin{itemize}
    \item approximate literals: 140 / counter literals: 216 = 64.8\%
    \item solution coverage: 64.9\%
    \item efficiency: 1.0
\end{itemize}

From the graph in totality, we can expect high efficiency when the $k$ value increases, but larger target variables depress it.

Focusing on the around approximate-at-most-25 of 30, the gray line, efficiencies indicate high at 24 and 26 but relatively low at 25.
The models at 24 and 26 are based on $h_1=2,w_1=4,h_2=2,w_2=2,m=2$, and 25 is based on $h_1=2,w_1=3,h_2=2,w_2=3,m=2$.
The former model seems efficient but approximate-at-most-25 cannot be implemented with the model.
More specifically, the model converts at-most-6 on the top into approximate-at-most-24 of 32 on the bottom (the case of 24), and at-most-7 on the top into approximate-at-most-28 of 32 on the bottom (the case of 26).
Since fixing the bottom variables to false decreases target variables and fixing to true decreases both count number and target variables, approximate-at-most-25 of 30 is not able to generate by the former efficient model with any adjustments of fixing target variables.
This makes approximate-at-most-25 relatively low efficiency.

\section{Discussion}
We studied the coverage which indicates how far approximate-at-most-k covers the possible solutions.
There are two types of coverage: overall coverage and maximum-count coverage; for example, about at-most-8 of 16 variables, considering overall 0$\sim$8 counts of true and considering only 8 count of true, respectively.
In this paper, we have mainly focused on the former, because it aims at the entire solution space and seems to be a comprehensive notion.
However, using an at-most-k constraint as a soft constraint, we generally want to find the maximum-count true case, such as 8 count of true for at-most-8.
Thus the maximum-count coverage will be considered for investigation in future work.

If an approximate-at-most-k covers 50\% of the possible solutions, then every possible solution is included in the approximate-at-most-k solutions with a probability of 50\%.
To solve a real-life problem, there is usually a huge space of possible solutions, and assuming 10 solutions there at this time, we can find the solution with a probability of 99.9\% ($1-0.5^{10}$).
From this point of view, we can expect practical utility more than the percentage of the coverage, and require a quantitative evaluation.

\section{Conclusion}

This paper proposes a new method for efficiently encoding at-most-k constraints into SAT, called approximate-at-most-k, which consumes fewer Boolean expressions than conventional encodings.
Approximate-at-most-k has gained low consumption at the cost of neglecting the completeness of the solution space, and it cannot be used to determine satisfiability.
Meanwhile, it is useful to search for better solutions together with soft constraints, in fewer Boolean expressions.

The experimental results support that approximate-at-most-k consumes relatively less than conventional counter encoding.
Considering the coverage of solution space, we observed the relationship between the reduction rate of literals and the coverage; for example, approximate-at-most-16 of 32 consumes only 15\% of counter encoding on the literal number and covers 44\% of the solution space.
In solving a real-world problem, approximate-at-most-k should be considered when there are massive soft constraints without sufficient computational resources.

\bibliography{the}

\end{document}